\documentclass[12pt]{article}
\usepackage{graphics}
\usepackage{amsmath}
\usepackage{fancyhdr}
\usepackage{amssymb}
\usepackage{latexsym}
\usepackage{graphicx}
\usepackage[hypertex]{hyperref}
\usepackage[authoryear]{natbib}
\hypersetup{backref, colorlinks=true}

\def\rz{\ifmmode{I\hskip -3pt R}
    \else{\hbox{$I\hskip -3pt R$}}\fi}
\def\nz{\ifmmode{I\hskip -3pt N}
    \else{\hbox{$I\hskip -3pt N$}}\fi}
\def\gz{\ifmmode{Z\hskip -4.8pt Z}
    \else{\hbox{$Z\hskip -4.8pt Z$}}\fi}
\def\cz{\ifmmode{C\hskip -4.8pt\vrule height5.8pt\hskip6.3pt}
\else{\hbox{$C\hskip -4.8pt\vrule height6.0pt\hskip6.3pt$}}\fi}
\def\qz{\ifmmode{Q\hskip -5.0pt\vrule height6.0pt depth0pt
   \hskip6pt}
    \else{\hbox{$Q\hskip -5.0pt\vrule height6.0pt depth0pt
   \hskip6pt$}}\fi}

\newcommand{\beq}{\begin{equation}}
\newcommand{\eeq}{\end{equation}}
\newcommand{\beql}{\begin{equation} \label}
\newcommand{\beqs}{\begin{eqnarray}}
\newcommand{\eeqs}{\end{eqnarray}}
\newcommand{\beas}{\begin{eqnarray*}}
\newcommand{\eeas}{\end{eqnarray*}}
\newcommand{\ber}{\begin{array}}
\newcommand{\eer}{\end{array}}
\newcommand{\becs}{\begin{cases}}
\newcommand{\eecs}{\end{cases}}

\newcommand{\leftm}{\left[\begin{array}}
\newcommand{\rightm}{\end{array}\right]}

\newcommand{\bfa}{{\bf a}}

\newcommand{\bfe}{{\bf e}}

\newcommand{\bfk}{{\bf k}}

\newcommand{\bfx}{{\bf x}}
\newcommand{\bfy}{{\bf y}}

\newcommand{\bfB}{{\bf B}}
\newcommand{\bfC}{{\bf C}}

\newcommand{\bfF}{{\bf F}}

\newcommand{\bfH}{{\bf H}}
\newcommand{\bfI}{{\bf I}}

\newcommand{\bfR}{{\bf R}}
\newcommand{\bfS}{{\bf S}}

\newcommand{\calE}{{\cal E}}
\newcommand{\calF}{{\cal F}}
\newcommand{\calG}{{\cal G}}

\newcommand{\calL}{{\cal L}}

\newcommand{\calU}{{\cal U}}

\newcommand{\calY}{{\cal Y}}

\newcommand{\Real}{{{\rm Re}}}
\newcommand{\Imag}{{{\rm Im}}}

\newcommand{\bfFhat}{{\hat{\bf F}}}

\newcommand{\bfxhat}{\hat{\bf x}}

\newcommand{\bfehat}{{\hat{\bfe}}}

\newcommand{\intbar}{{\int\!\!\!\!\!- }} 
\newcommand{\inttbar}{{\int\!\!\!\!\!\!- }} 

\newcommand{\eps}{{\varepsilon}}

\newcommand{\half}{\frac{\;1}{\;2}}

\newcommand{\kk}{ {k}}

\newcommand{\aand}{{{\rm\;\; and\;\;}}}

\newcommand{\iif}{{{\rm if}}}

\newcommand{\oon}{{{\rm \;on\;}}}

\newcommand{\aas}{{{\rm as}}}

\newcommand{\oor}{{{\rm or}}}

\newcommand{\deltatld}{{\tilde{\delta}}}
\newcommand{\Yp}{{Y}}
\newcommand{\bfxtld}{{\tilde{\bfx}}}

\newcommand{\uc}{{u_c}}
\newcommand{\phic}{{\phi_c}}
\newcommand{\uczero}{{u_c^0}}

\newcommand{\phiczero}{{\phi_c^0}}
\newcommand{\ucone}{{u_c^1}}
\newcommand{\phicone}{{\phi_c^1}}

\newcommand{\Diverg}{{\rm Div}}

\begin{document}

\begin{center}
{\huge An analysis of the field theoretic approach to the quasi-continuum method}\\

\vspace{0.5cm} {{Vikram Gavini$^{a,*}$} and Liping Liu$^{b}$}\\

\vspace{0.5cm} $^a$Department of Mechanical Engineering, University of Michigan, Ann Arbor, MI 48109-2125, USA\\
$^b$Department of Mechanical Engineering, University of Houston, Houston, TX 77204-4006 USA\\
$^*$Corresponding Author (vikramg@umich.edu)
\end{center}

\begin{abstract}

Using the orbital-free density functional theory as a model theory,
we  present an analysis of the field theoretic approach to quasi-continuum method.
In particular, by perturbation method and multiple scale analysis, we provide a
formal justification for the validity of the coarse-graining of
various fields, which is central to the quasi-continuum reduction of field
theories. Further, we derive the homogenized equations that govern
the behavior of electronic fields in regions of smooth deformations.
Using Fourier analysis, we determine the far-field solutions for
these fields in the presence of local defects, and subsequently
estimate cell-size effects in computed defect energies.
\end{abstract}

\section{Introduction}
The quasi-continuum method has, in the past decade, become an
important  computational technique to study the behavior of defects
in materials where a wide range of interacting length scales become
important. The main idea behind the quasi-continuum method is a
seamless bridging between the various length scales of interest by
imposing kinematic constraints on the degrees of freedom and
systematically coarse-graining away from the regions of interest.
The quasi-continuum method was first proposed in the context of
empirical interatomic potentials \citep{TOP1996}, where the energy
of the system was expressed as a non-local sum over the positions of
atoms. The kinematic constraints on the positions of atoms---degrees
of freedom in the formulation---are imposed via an unstructured
finite-element triangulation of atomic positions with full atomistic
resolution in regions of interest, for instance at the core of a
defect, and rapidly coarse-grains away to capture the long-range
elastic effects. Apart from the kinematic
constraints introduced on the degrees of freedom, further
approximations are introduced to reduce the computational complexity
of the formulation. The differing nature of these approximations,
which include invoking the Cauchy-Born hypothesis in some regions of
the model or introducing cluster summation rules in the spirit of
numerical quadratures, have resulted in many different formulations of
the quasi-continuum method. We refer to
\citet{SMTRPO1999,KO2001,TM2002,SMSJ2004,ES2009} and reference
therein for a comprehensive overview of the different formulations of the
method. Recent investigations and numerical analysis of the method
\citep{SMSJ2004,ELY2006,DL2008,LO2009,DLO2009} suggest that these
approximations can result in undesirable consequences, namely, lack of a
variational structure, lack of stability and consistency of the
approximation schemes, and uncontrolled errors in some cases.

In a recent work \citep{GBO2007} the quasi-continuum method was
developed for electronic structure calculations using orbital-free
density functional theory (OFDFT). OFDFT, which is an approximation to the
widely used Kohn-Sham formulation of density functional theory
\citep{HK1964,KS1965}, describes the ground-state energy of
the system as an explicit functional of electron-density and is
valid in materials systems whose electronic structure is close to a
free electron gas (cf. \citet{Parr,WT1992,SM1994,WGC1998,WGC1999}
for a comprehensive overview). The quasi-continuum reduction of
OFDFT was achieved using a real-space local variational formulation, and a coarse-graining of the electronic-fields and
positions of atoms---degrees of freedom in the formulation---through kinematic constraints
imposed using nested finite-element triangulations. An
important difference in the mathematical structure of
quasi-continuum formulation for OFDFT in comparison to empirical
interatomic potentials is that OFDFT is a local field theory as
opposed to the non-local description of extended interactions in
empirical potentials. A local field formulation, as in the case of
OFDFT, admits quadrature approximations to further reduce the
computational complexity without introducing the undesirable
consequences characteristic of conventional quasi-continuum formulations.

In the prequel to this article \citep{IG2010}, we suggest a field
formulation for commonly used interatomic potentials, where the
extended interactions in these potentials are reformulated into a
local form by constructing partial differential equations (PDE's) whose Green's functions correspond
to the kernels of the non-local interactions. We further demonstrate
that the quasi-continuum reduction of these field formulations is
variational, a consistent numerical approximation, and provides
significantly better accuracy than previous
formulations. Moreover, the field formulation of interatomic
potentials provides a unified framework where the quasi-continuum
reduction is solely a numerical approximation scheme irrespective of
the field theory used to describe the system---density functional
theory or field theories that represent interatomic interactions.

In the quasi-continuum reduction of field theories
\citep{IG2010,GBO2007}, the various fields that appear in the
formulation are decomposed into predictor fields and corrector
fields. The predictor fields are computed by performing a periodic
calculation using the Cauchy-Born hypothesis, and the corrector fields are subsequently
computed from the variational formulation. For smooth deformations
which do not depend on the atomic-scale,
\citet{BLL2002} show that the various fields are given, to the leading order, by a
periodic calculation using the Cauchy-Born hypothesis. Hence, in regions
away from the defect-core it is expected that the predictor
fields are good approximations to the fields. Thus,
the corrector fields are represented on a finite-element
triangulation which is subatomic near the core and coarse-grains
away to become superatomic, and this constitutes the quasi-continuum
reduction of field formulations.

The representation of the corrector fields on a coarse-grained
triangulation  is valid under the hypothesis that corrector fields
do not exhibit oscillations on the atomic-scale. In this work, we
provide a formal justification for this hypothesis. We conduct our
analysis in the framework of OFDFT and latter comment on other field
theories. We first use the perturbation method to find the governing equations for the
corrector fields corresponding to a weak defect. While the defect
plays the role of a source (forcing function), the coefficients of these governing
equations are given by the unperturbed (predictor) electronic fields and hence oscillate on
the atomic-scale. For homogeneous deformation, the unperturbed electronic fields  are given by periodic
calculations with respect to the unit cell at atomistic scale; for a smooth macroscopic deformation, the
unperturbed electronic fields are generally unknown. Motivated by the
thermodynamic limit \citep{BLL2002,GLE2007}, we nevertheless hypothesize
the unperturbed fields are given by periodic
calculations with respect to the local atomistic lattice.
Further, since the unperturbed electronic fields oscillate at the atomistic scale which is much
smaller than the macroscopic supercell and the considered macroscopic perturbation, we
employ the multi-scale analysis \citep{Cioranescu1999} to find the corrector electronic fields.
In particular, we demonstrate that the corrector electronic fields to their leading order and
first order are independent of the lattice parameter, and hence do
not exhibit atomic-scale oscillations. However, this result shall be
interpreted with caution near the defect-core since in reality a
defect, e.g., a vacancy or an interstitial, is localized at the
atomic-scale.  Further, we derive the homogenized equations for the
corrector fields. These homogenized equations turn out to be a
second-order linear system of PDE's. By Fourier method, we find
their Green's functions explicitly, which show that the correctors
fields in OFDFT, corresponding to electrostatic potential and
electron density, decay exponentially. Additionaly, we compute
their solutions for a situation representative of a vacancy in an
infinite crystal. Using these solutions, we analyze the cell-size
effects arising from a computation on a finite domain and estimate
the domain size required for achieving chemical accuracy in vacancy
formation energy. Our results show that a cell-size of the order of
1,000 atoms is required to attain a converged value for the vacancy
formation energy in aluminum, which is much larger than the
cell-sizes that are  commonly used in numerical simulations. This
estimate is in agreement with a recent cell-size study in
\citet{GBO2007}. We further note that the cell-size effects are
likely to be more significant for defects like dislocations where
the decay in elastic fields is much slower.

The remainder of this paper is organized as follows. In
section~\ref{sec:setup} we formulate OFDFT with
Thomas-Fermi-Weizsacker kinetic energy functionals, and present the
problem definition and the assumptions made in this analysis. In
section~\ref{sec:perturbation} we discuss the perturbation analysis
of the corrector fields, and present the multi-scale analysis of
these fields and derive the homogenized equations in
section~\ref{sec:homogenization}. In sections~\ref{sec:farFields}
and \ref{sec:defectCell-size} we derive the Green's functions of the
homogenized equations and compute their solutions for a spherical
defect representing a vacancy. In section~\ref{sec:extension} we
comment on the extension of this analysis to  other flavors of OFDFT
which use non-local kernel energies and other field formulations
representing empirical interatomic potentials. We finally conclude
in section~\ref{sec:conclusions} providing an outlook.

\section{Problem definition}\label{sec:setup}
Consider an infinite crystal with lattice points given by
$\calL_\eta=\eta \calL$ and
$\calL=\{\sum_{i=1}^3\nu_i\bfehat_i:\nu_1,\nu_2,\nu_3\in \gz\}$,
where $\eta<<1$ denotes the fixed lattice parameter and
$\bfehat_1,\bfehat_2,\bfehat_3\in \rz^3$ are the rescaled lattice
vectors satisfying $\bfehat_3\cdot (\bfehat_1\times \bfehat_2)=1$.
We refer to
 \beas  U_\eta=\eta U_0,\qquad U_0=\{\sum_{i=1}^3 t_i\bfehat_i: \;-\half
< t_1,t_2,t_3< \half\} \eeas as the unit cell and the rescaled unit
cell, respectively. Let $Y_0=(-1,1)^3$ be a macroscopic supercell such that it overlaps with an integer number
of unit cells, $Z$ be the charge at each nucleus measured in units
of electron charge, and $\bfy:Y_0\to \Yp$  be a smooth macroscopic
deformation that carries a reference point $\bfx_0\in Y_0$ to a new
point $\bfy(\bfx_0)\in \Yp$. In this work we are interested in
macroscopic deformations that are independent of $\eta$. We assume
that the nuclei follow the Cauchy-Born rule, and hence the nuclear
charges in the deformed configuration are given by \beas
b_{\bfy}(\bfx)=\sum_{\bfx_0\in \calL_\eta\cap Y_0}Z
\deltatld(\bfx-\bfy(\bfx_0)), \eeas where $\deltatld$ is a
regularization of the Dirac distribution that represents a unit
nuclear charge.

To present our ideas we consider the energy of a system described by OFDFT.
We remark that the ideas presented
here are general and can be equally applied to any field theory, for
instance, fields theories that describe empirical interatomic
potentials discussed in \citet{IG2010}. In density functional
theory, the energy of a material system is given by \beqs
    E(u, b_{\bfy})
    =
    T_s(u)+E_{xc}(u)+E_H(u)+E_{ext}(u,
    b_{\bfy})+E_{zz}(b_{\bfy})\,,
\eeqs where $u$ denotes the square-root electron density, $T_s$
denotes the kinetic  energy of non-interacting electrons, $E_{xc}$
denotes the exchange and correlation energies that account for the
quantum mechanical effects, and $E_H$, $E_{ext}$, $E_{zz}$ denote
classical electrostatic interaction energies between electrons and
nuclei. In OFDFT, $T_s$ is approximated by explicit functional forms of electron density
as opposed to the Kohn-Sham approach where it is computed exactly
within the mean field approximation. A simple choice for this
approximation is the Thomas-Fermi-Weizsacker (TFW) family of kinetic
energy functionals \citep{Parr}: \beqs\label{eq:TFW}
    {T}_{s}(u)
    =
    C_F\int_\Yp{{u}^{10/3}d\bfx}
   +
    \frac{\lambda}{2}
    \int_\Yp{{|\nabla u|}^2}
    d\bfx,
\eeqs where $0\leq\lambda\leq1$ is a parameter and $C_F=\frac{3}{10}(3\pi^2
)^{2/3}$. More  accurate kinetic energy functionals have been
proposed in the past decade which account for the linear response
of a uniform electron gas. For clarity we postpone our analysis of
these functionals to section~\ref{sec:extension}. By choosing the
Thomas-Fermi-Weizsacker functionals~\eqref{eq:TFW} for kinetic energy, and
following the real-space formulation of OFDFT proposed in
\citet{GKBO2007}, we express the total energy of the system as \beqs
\label{eq:Eb} E(u,\phi;b_{\bfy})=  \int_{\Yp }
\Big[f(u)+\frac{\lambda}{2} |\nabla u|^2 -\frac{1}{2} |\nabla
\phi|^2+(u^2+b_{\bfy})\phi \Big]d\bfx, \eeqs where $f(u)=C_F
u^{10/3}$ and $\phi$ denotes the trial electrostatic potential.
In the above, we ignore
exchange and correlation energies and comment on them in
section~\ref{sec:extension}. The ground state of $(\phi, u,\bfy)$ is
determined by the following min-max problem \beqs \label{eq:calE}
\calE_{tot}(0):= \min_{\bfy\in \calY }\Big\{ \calE(\bfy):=\min_{u\in
\calU(b_{\bfy})}\max_{\phi\in H^1_{per}(\Yp)  } E(u,\phi;b_\bfy)
\Big\}, \eeqs
 where
 \beqs \label{eq:calAb}
\calY:&=&\{\bfy\in \chi:\nabla \bfy \mbox{\,\,periodic in\,\,} Y_0\},  \nonumber \\
\calU(b_{\bfy}):&=&\{u\in H_{per}^1(\Yp ):\;\int_\Yp
(u^2+b_{\bfy})d\bfx=0,\;u\geq 0\}.
 \eeqs
  In the above definitions, $\chi$ is a suitable function space that admits
minimizers of $\calE(\bfy)$. In this analysis, since our focus is to
derive and analyze the far-field behavior of the displacement and
electronic fields, we restrict our attention to a local minimizer
of $\calE(\bfy)$ in $\calY$.

Let $(\phi_\bfy, u_\bfy)$ be a solution of the min-max problem for a
smooth deformation $\bfy\in \calY$
 \beqs \label{eq:phiyuy}
 \calE(\bfy)=E(u_\bfy,\phi_\bfy;b_\bfy)=\min_{u\in
\calU(b_\bfy)}\max_{\phi\in H^1_{per}(\Yp)  } E(u,\phi;b_\bfy).
\eeqs The existence of a solution for the saddle point
problem~\eqref{eq:phiyuy} can be established following the ideas in
\citet{GKBO2007}, where the analysis was carried out in a
non-periodic setting with Dirichlet boundary conditions on a bounded
domain. We remark that the arguments in \citet{GKBO2007} can be
appropriately modified to the periodic setting, and these details
are not discussed in this article to maintain our focus on
multi-scale analysis.
 We also refer to \citet{Lieb1981} for results
on the existence and uniqueness of solutions for various flavors of
OFDFT.

It is clear from the definition~\eqref{eq:Eb}, if $( \phi_\bfy,
u_\bfy)$ is a solution to the min-max problems in
equation~\eqref{eq:phiyuy}, so is $(\phi_\bfy+c, u_\bfy)$ for any
$c\in \rz$. By the standard first-variation  calculations it follows
that there exists a solution to the min-max problem in
equation~\eqref{eq:phiyuy}, denoted by $(\phi_\bfy,u_\bfy)$, satisfying
 \beqs
\label{eq:ELy} \becs
\Delta \phi_\bfy+   (u_\bfy^2+b_\bfy)=0&\oon \;\Yp ,\\
-\lambda \Delta  u_\bfy +f'(u_\bfy) +2u_\bfy\phi_\bfy=0 &\oon\;\Yp ,\\
 \mbox{subject to:}\quad  u_\bfy\in \calU(b_\bfy);\quad  \phi_\bfy\in H^1_{per}(\Yp).&\;\; \\
\eecs \eeqs Note that, in the above equation, the Lagrangian
multiplier associated  with the constraint in
equation~$\eqref{eq:calAb}_2$ has been absorbed  into the
electrostatic potential $\phi_\bfy$. Thus, the solution~$\phi_\bfy$
to problem \eqref{eq:ELy} no longer allows an arbitrary additive
constant (cf. \citet{CLL1998} for further discussion on this point).

We now discuss the nature of the solution $(\phi_\bfy, u_\bfy)$ to
problem~\eqref{eq:ELy}. First we assume a homogeneous deformation with
$\nabla \bfy=\bfF_0\in \rz^{3\times 3}$ on $Y_0$. Consider problem \eqref{eq:ELy}
on the deformed unit cell $\bfF_0U_\eta$ \beqs
\label{eq:ELyunit}
&&\becs
\Delta \phi+   (u^2+b_\bfy)=0&\oon \;\bfF_0 U_\eta ,\\
-\lambda \Delta  u +f'(u)
 +2u\phi =0 &\oon\;\bfF_0 U_\eta,\\
 \mbox{subject to:\;\; }\int_{\bfF_0 U_\eta} (u^2+b_\bfy)d\bfx=0,\quad  u,\phi\in H_{per}^1(\bfF_0 U_\eta).
\eecs
\eeqs
For $f(u)=\frac{3}{10}(3\pi^2 )^{2/3} u^{10/3}$,
\citet{CLL1998} have shown
 that the periodic extension of the solution to problem \eqref{eq:ELyunit} with respect to
 the period $\bfF_0U_\eta$,   denoted by $(\phi^\ast, u^\ast)$,
is the solution to problem \eqref{eq:ELy}:
\beqs \label{eq:uphiunitcell0}
\phi_\bfy(\bfx)=\phi^\ast( \bfx), \qquad u_\bfy(\bfx)=u^\ast( \bfx) \qquad \forall\,\bfx\in \bfF_0Y_0.
 \eeqs
For future convenience, we denote by
 \beqs \label{eq:uphiunitcel00}
\phi_p( \bfF_0, \bfx)=\phi^\ast(\eta \bfx), \qquad u_p(\bfF_0, \bfx)=u^\ast(\eta \bfx),
 \eeqs
 where the subscript $_p$ signifies that  $\bfx\mapsto (\phi_p(\bfF_0,\bfx),
u_p(\bfF_0,\bfx))$  are periodic with period equal to the rescaled
unit cell~$\bfF_0 U_0$. It is worthwhile noticing that $(\phi_p,
u_p)$ are  considered as being defined by the exact solutions to the
unit cell problem~\eqref{eq:ELyunit} through equation
\eqref{eq:uphiunitcel00}, instead of the solutions in the
thermodynamic limit discussed in \citet{BLL2002}.
 By
equations~\eqref{eq:uphiunitcell0} and \eqref{eq:uphiunitcel00}, we
have \beqs \label{eq:phi0u0}
 \phi_\bfy(\bfx)=\phi_p( \bfF_0, \bfxtld), \qquad u_\bfy(\bfx)=u_p(\bfF_0, \bfxtld),
\eeqs
 where  $\bfxtld=\bfx/\eta$ denotes the fast variable in the subsequent homogenization calculation.
We remark that, in Section 4,
equations~\eqref{eq:phi0u0} and  the fact that $\eta<<1$ compared
with the macroscopic supercell $Y_0$ will be used  to derive  the
homogenized equations for the corrector fields. This homogenization
limit is not the thermodynamic limit where the nuclei are assumed to
locate at $\eta \calL\cap Y_0$ and $\eta\to 0$. Trying to couple the
homogenization limit and the thermodynamic limit encounters
difficulties, and we will not address this issue in this paper.
Further, we identity the solution to problem \eqref{eq:ELyunit}
 is also a solution of the min-max problem \beqs
\label{eq:phiyuyunitcell} W(\bfF_0):=\min_{u}
\max_{\phi}\inttbar_{\bfF_0 U_\eta } \Big[f(u)+\frac{\lambda}{2}
|\nabla u|^2 -\frac{1}{2} |\nabla \phi|^2+(u^2+b_\bfy)\phi
\Big]d\bfx \eeqs subject to the same constraints as in
equation~\eqref{eq:ELyunit}. Here and subsequently, $\intbar_\Omega
(\;)=\frac{1}{volume(\Omega)} \int_\Omega(\;)$ denotes the averaged
value of the integrand over the domain $\Omega$.
 In terms of the solutions $(\phi_p, u_p)$ to the unit cell problem, we define
the following  quantities for future use \beqs \label{eq:alphabeta}
&&\alpha(\bfF_0)=\inttbar_{\bfF_0U_0} u_p(\bfF_0, \bfxtld) d\bfxtld
,\qquad\beta(\bfF_0)=\inttbar_{\bfF_0U_0}
\phi_p(\bfF_0, \bfxtld) d\bfxtld, \nonumber\\
&&\gamma(\bfF_0)=\inttbar_{\bfF_0U_0} [\half
f^{''}(u_p(\bfF_0,\bfxtld) )] d\bfxtld +\beta(\bfF_0)\, .\eeqs
As in classical continuum mechanics, all the functions $W, \alpha,
\beta, \gamma:\rz^{3\times 3}\to \rz$ satisfy the material frame
indifference and material symmetries \beqs \label{eq:RH}
&&W(\bfR \bfF_0 )=W(\bfF_0)\qquad \forall\,\bfR\in SO(3)\; \&\; \bfF_0\in  \rz^{3\times 3}\; \mbox{with}\; det {\bfF}_0>0,\nonumber \\
&&W( \bfF_0\bfH )=W(\bfF_0)\qquad \forall\,\bfH\in  \calG(\bfF_0),
\eeqs where $SO(3)$ consists of all rigid rotation matrices,
$\calG(\bfF_0)$ is the point group associated with the Bravais
lattice $\calF_0\calL$, and $W$ can be replaced by $\alpha,
\;\beta,$ $\oor\;\gamma$ in equation~\eqref{eq:RH}.
Equation~\eqref{eq:RH} can be verified directly from the
definitions~\eqref{eq:phiyuyunitcell} and~\eqref{eq:alphabeta}.

We now consider the case when the deformation $\bfy$ has a smooth
macroscopic  deformation gradient  $\bfFhat:\Yp\to \rz^{3\times 3}$
on the current configuration \beqs \label{eq:bfFtld}
\bfFhat(\bfx)=\bfF(\bfy^{-1}(\bfx)) \;\;\forall\;\bfx\in \Yp,\qquad
\bfF(\bfx_0)=\nabla \bfy(\bfx_0)\;\;\forall\;\bfx_0\in Y_0. \eeqs
A priori, for this case, we have no knowledge on  the solution to \eqref{eq:ELy}.
Since $\eta<<1$, motivated by  \citet{BLL2002} we hypothesize that the solution to \eqref{eq:ELy} is
 given by
 \beqs \label{eq:phiulimits}
\phi_\bfy(\bfx) =\phi_p(\bfFhat(\bfx), \bfxtld), \qquad
u_\bfy(\bfx)= u_p(\bfFhat(\bfx), \bfxtld),
 \eeqs
and the elastic energy is  given by \beqs \label{eq:calEW}
\calE(\bfy)=E(u_\bfy,\phi_\bfy;b_\bfy)= \int_\Yp
W(\bfFhat(\bfx)) d\bfx, \eeqs
 where $(\phi_p, u_p)$ are defined by the exact solutions to the unit cell problem~\eqref{eq:ELyunit}
through equation~\eqref{eq:uphiunitcel00}, and $W:\rz^{3\times 3}\to \rz$ is
the elastic energy density on the deformed configuration $\Yp$ given
by equation~\eqref{eq:phiyuyunitcell}.

The solution to the outer minimization problem~\eqref{eq:calE} may not be unique, and throughout this work we will restrict our attention to local minimizers
that satisfy the Euler-Lagrange equation corresponding to the energy in equation~\eqref{eq:calEW}, which is the familiar equilibrium equation of elasticity \beqs
\label{eq:ELelastic} \Diverg\,\bfS(\nabla \bfy^\ast)=0 \qquad
\oon\;Y_0, \eeqs where \beas
 [\bfS(\nabla \bfy^\ast)]_{pi}=\frac{\partial JW}{\partial \bfF_{pi}}(\nabla \bfy^\ast), \qquad J(\bfF)=\det(\bfF).
\eeas
Note that in equation~\eqref{eq:ELelastic}, $\bfS$ is the first
Piola-Kirchhoff stress and ${\rm Div\,\bfS}=\frac{\partial
\bfS_{pi}}{\partial {\bfx_0}_i}$.

\section{Perturbation analysis}\label{sec:perturbation}
We now consider the effect of defects on electronic fields $(\phi,
u)$. A defect  breaks the lattice symmetry which in effect is a
perturbation of the forcing term,
$b_\bfy$, in equation \eqref{eq:ELy}. Thus, we replace the forcing term in equation
\eqref{eq:ELy} by a small perturbation of $b_\bfy$:
$b_\bfy^\eps=b_\bfy+\eps b_c$ with $\eps<<1$ and consider $b_c$ to be independent of
the lattice parameter $\eta$ which allows us to subsequently pass to the homogenization
limit in section~\ref{sec:homogenization}. If $b_c$ has a compact
support, this perturbation can be interpreted as a weak local defect, formed by slowly
reducing the charges on the nuclei in a macroscopic region, in an otherwise
perfect crystal undergoing a smooth deformation. We
are interested in calculating the influence of this perturbation
(defect) on the ground state of OFDFT
 and, in particular, on the total energy. As in equation
\eqref{eq:calE}, the ground state of the system is governed by
 \beqs
\label{eq:minmaxeps} \calE_{tot}(b_c):=\min_{\bfy\in
\calY}\Big\{\calE^\eps(\bfy;b_c):=\min_{u\in
\calU(b^\eps_\bfy)}\max_{\phi\in H^1_{per}(\Yp)  }
E(u,\phi;b_\bfy+\eps b_c) \Big\}.
 \eeqs
Note that if $b_c=0$, i.e., the system is unperturbed, then
$\calE_{tot}(b_c=0)$ is equal to $\calE_{tot}(0)$
in equation~\eqref{eq:calE}.

We solve the above problem approximately by perturbation method. We
first consider the inner min-max problem in equation
\eqref{eq:minmaxeps} for given $\bfy\in \calY$. Let \beqs
\label{eq:uphieps} \phi^\eps=\phi_\bfy+\eps\phic\in H^1_{per}(\Yp) ,
\qquad u^\eps=u_\bfy+\eps {\uc}  \in \calU(b_\bfy^\eps), \eeqs be
the solutions, where $(\phi_\bfy,u_\bfy)$,  the
solutions to the  unperturbed problem~\eqref{eq:ELy},  are
referred to as the predictor fields in the quasi-continuum
formulation,  and $(\phi_c,u_c)$ are referred to as  the corrector fields
\citep{GBO2007}.
Inserting equation~\eqref{eq:uphieps} into equation~$\eqref{eq:calAb}_2$,
we obtain the charge neutrality constraint
 \beqs \label{eq:uepsC}
\int_\Yp  [2u_\bfy {\uc} +\eps u_c^2 +b_c]d\bfx=0. \eeqs
Inserting equation \eqref{eq:uphieps} into equation \eqref{eq:Eb}, we expand the energy as \beqs
\label{eq:Eeps} &&E(u^\eps,\phi^\eps;b_\bfy+\eps
b_c)=E(u_\bfy,\phi_\bfy;b_\bfy)+\eps\int_{\Yp } \Big\{
\big[f'(u_\bfy) -\lambda \Delta  u_\bfy+2u_\bfy \phi_\bfy \big]
{\uc}
\nonumber\\
&&\hspace{4cm} +[\Delta \phi_\bfy +u_\bfy^2+b_\bfy]\phic+ b_c \phi_\bfy \Big\}d\bfx + \eps^2E_2({\uc} ,\phic, \bfy; b_c)+o(\eps^2) \nonumber\\
&&\hspace{3.3cm} =E(u_\bfy,\phi_\bfy;b_\bfy)+\eps\int_Yb_c\phi_\bfy
d\bfx+\eps^2 E_2({\uc} ,\phic, \bfy; b_c) +o(\eps^2), \eeqs where
the second equality follows from the Euler-Lagrange equations in~\eqref{eq:ELy} for
$(\phi_{\bfy},u_{\bfy})$, and
\beqs \label{eq:E2} E_2({\uc} ,\phic, \bfy; b_c)=\int_{\Yp }
\Big[\half f^{''}(u_\bfy) {\uc} ^2 +\frac{\lambda}{2} |\nabla {\uc}
|^2  -\frac{1}{2 } |\nabla \phic|^2 +(2u_\bfy \uc +b_c)\phic
+\phi_\bfy u_c^2\Big]d\bfx. \eeqs
Neglecting $o(\eps^2)$-terms in equation~\eqref{eq:Eeps},
by the  inner min-max problem in equation~\eqref{eq:minmaxeps} we
arrive at the following min-max problem for
$(\uc,\phic)$:
 \beqs \label{eq:minmaxc} \calE_2(\bfy;b_c):=
\min_{{\uc} }\max_{\phic  } E_2({\uc} ,\phic,\bfy;b_c) \,\eeqs
subject to the constraints (cf. equation~\eqref{eq:uepsC}) \beqs
\label{eq:calAtld} \phic \in H^1_{per}(\Yp),\quad
 \uc\in H_{per}^1(\Yp), \quad \int_\Yp  (2u_\bfy u_c+b_c)d\bfx=0.
\eeqs We remark that the zeroth and first order terms in
equation~\eqref{eq:Eeps} are absent in the min-max
problem~\eqref{eq:minmaxc} since they are independent of
($\phic,{\uc} $). By the standard first-variation calculations, we
show that the  Euler-Lagrange equations for $(\phic, {\uc} )$
associated with the min-max problem~\eqref{eq:minmaxc} are \beqs
\label{eq:ELyc} \becs
\Delta \phic+   (2u_\bfy {\uc}+b_c)=0 &\oon\;\Yp,\\
-\lambda \Delta {\uc} +(f^{''}(u_\bfy)+2\phi_\bfy){\uc} +2 u_\bfy \phic=0 &\oon\;\Yp,\\
\eecs \eeqs where, as in equation \eqref{eq:ELy}, we have absorbed  into the
potential $\phi_c$ the Lagrangian multiplier associated with the last
constraint in equation \eqref{eq:calAtld}, which is a constant independent of
$\bfx$. We further notice  that equations~\eqref{eq:ELyc} can be
obtained by linearizing equation~\eqref{eq:ELy} near the  solutions
$(\phi_\bfy, u_\bfy)$. We remark that although the perturbation
analysis was conducted under the assumption of weak local defects,
the perturbation expansion given by equation~\eqref{eq:uphieps} is a
reasonable assumption in regions away from defects that are not
necessarily weak. This follows as the perturbations in electronic fields decay away from the defect core due to the elliptic nature of the PDE's, and the governing equations for corrector fields
will subsequently be valid in these regions.

\section{Homogenization}\label{sec:homogenization}

We now turn towards establishing certain properties of the corrector
fields which play a fundamental role in the construction of
quasi-continuum reduction of field formulations proposed in
\citet{GBO2007,IG2010}, and provide a formal mathematical
justification for the method. Before proceeding to details, we notice
the following useful identity. Let $f(\bfx,\bfxtld)$ be a smooth
function which is periodic in the second variable $\bfxtld$ with
period $\bfFhat(\bfx)U_0$. If $\eta<<1$, we have the identity (cf.
e.~g. \citet{Cioranescu1999}, Chapter 2), \beqs \label{eq:f2sint} \int_\Yp
f(\bfx,\frac{\bfx}{\eta})d\bfx=\int_\Yp \inttbar_{\bfFhat(\bfx)U_0}
f(\bfx,\bfxtld)d\bfxtld d\bfx+o(1). \eeqs

Since the unperturbed solutions $(\phi_\bfy, u_\bfy)$ given by
equation~\eqref{eq:phiulimits} oscillate at the atomic scale-$\eta$,
 presumably the corrector field solutions ($\phic,{\uc} $) to
the governing equations in~\eqref{eq:ELyc} oscillate at
the $\eta$-scale as well. In this section, we determine the order of this $\eta$-scale oscillation in the corrector fields
($\phic,{\uc} $) and
 whether this atomic-scale oscillation is important
to the leading order in energy. Further, we determine the
homogenized equations that govern the macroscopic behavior of these
corrector fields. To this end, following the method of the multiple
scale expansions, we assume
\beqs \label{eq:phiueta} \becs
\phic(\bfx)=\phiczero(\bfx,\bfxtld)+\eta \phicone (\bfx,\bfxtld)+\cdots, \\
{\uc} (\bfx)=\uczero (\bfx,\bfxtld)+\eta \ucone (\bfx,\bfxtld)+\cdots,\\
\eecs \eeqs where $\bfxtld=\bfx/\eta$ is the fast variable,
$\phi_c^i(\bfx,\bfxtld),\, u_c^i(\bfx,\bfxtld)$ ($i=0,1,\cdots$) are
assumed to be periodic  in the fast variable $\bfxtld$ with period
$\bfFhat(\bfx) U_0$. Replacing $(\phi_\bfy, u_\bfy)$ by the right hand side of
 equation~\eqref{eq:phiulimits},  we
rewrite $E_2$ in equation~\eqref{eq:E2} as \beqs \label{eq:E2Hom}
 E_2({\uc} ,\phic, \bfy; b_c)=\int_{\Yp } \Big[(\half f^{''}(u_p)+\phi_p)\uc^2
+\frac{\lambda}{2} |\nabla {\uc} |^2
 -\frac{1}{2 } |\nabla \phic|^2 +(2u_p \uc
+b_c)\phic \Big]d\bfx. \eeqs
Inserting the multiple scale expansion~\eqref{eq:phiueta}
into equation~\eqref{eq:E2Hom}, we have \beqs \label{eq:E2eta}
E_2({\uc} ,\phic,\bfy;  b_c)&=&\int_{\Yp } \Big[(\half
f^{''}(u_p)+\phi_p) (\uczero )^2 +\frac{\lambda}{2}
|\frac{1}{\eta}\nabla_\bfxtld  \uczero +\nabla_\bfx  \uczero
+\nabla_\bfxtld \ucone |^2 \nonumber \\
&&-\frac{1}{2 } |\frac{1}{\eta}\nabla_\bfxtld \phiczero+\nabla_\bfx
\phiczero+\nabla_\bfxtld \phicone |^2 +(2u_p
\uczero +b_c)\phiczero\Big]d\bfx+O(\eta) \nonumber \\
&=&\frac{1}{2\eta^2} \int_\Yp \Big[\lambda|\nabla_\bfxtld  \uczero |^2-|\nabla_\bfxtld  \phiczero|^2 \Big]d\bfx\nonumber\\
&&+\frac{1}{\eta} \int_\Yp \Big[\lambda\nabla_\bfxtld \uczero  \cdot
(\nabla_\bfx  \uczero +\nabla_\bfxtld \ucone )
-\nabla_\bfxtld  \phiczero \cdot (\nabla_\bfx  \phiczero+\nabla_\bfxtld \phicone ) \Big]d\bfx \nonumber \\
&&+\int_{\Yp } \Big[(\half f^{''}(u_p)+\phi_p)(\uczero
)^2 +\frac{\lambda}{2} |\nabla_\bfx  \uczero +\nabla_\bfxtld \ucone|^2  \\
&&\hspace{1.5cm} -\frac{1}{2 } |\nabla_\bfx \phiczero+\nabla_\bfxtld
\phicone |^2 +(2u_p \uczero
+b_c)\phiczero\Big]d\bfx+O(\eta).\nonumber \eeqs We neglect the
higher order terms of $O(\eta)$ in equation~\eqref{eq:E2eta}. Since
$\eta<<1$, we consider the min-max problem~\eqref{eq:minmaxc} first for the leading
$\frac{1}{\eta^2}$-terms in equation~\eqref{eq:E2eta}, which is
given by
 \beas \min_{\uczero}\max_{\phiczero  }
 \int_\Yp \Big[\lambda|\nabla_\bfxtld  \uczero |^2-
|\nabla_\bfxtld  \phiczero|^2\Big]d\bfx. \eeas It is clear that a
solution to the above problem necessarily satisfies
 \beqs \label{eq:uphixtld}
\nabla_\bfxtld  \uczero (\bfx,\bfxtld)=0\qquad \aand\qquad
\nabla_\bfxtld  \phiczero(\bfx,\bfxtld)=0,
 \eeqs
which means that $\phiczero(\bfx,\bfxtld)$ and $\uczero
(\bfx,\bfxtld)$ are independent of
 the fast variable $\bfxtld$ and hence can be rewritten as
 \beqs \label{eq:uphix}
 \uczero (\bfx,\bfxtld)=\uczero (\bfx)\qquad \aand\qquad  \phiczero(\bfx,\bfxtld)=\phiczero(\bfx).
 \eeqs
This shows that the leading order terms in the corrector fields do not
exhibit atomic-scale oscillations,  and thus the corrector fields can
be resolved accurately on length scales larger than the lattice
parameter. This key result formally justifies the
coarse-graining of corrector fields introduced in the quasi-continuum reduction of field
theories.

Further, in account of equation~\eqref{eq:uphixtld}, the
$\frac{1}{\eta}$-terms   on the right hand side of
equation~\eqref{eq:E2eta} vanish. Finally, we consider the
$\frac{1}{\eta^0}$-order terms on the right hand side of
equation~\eqref{eq:E2eta} which represent the leading order terms in
the multiple scale expansion of $E_2$. Using
equation~\eqref{eq:f2sint} we can rewrite equation~\eqref{eq:E2eta}
as \beqs \label{eq:E2eta0order} E_2({\uc} ,\phic,\bfy;  b_c)
&\approx& \int_{\Yp } \inttbar_{\bfFhat(\bfx) U_0}
\Big[\frac{\lambda}{2} |\nabla_\bfx  \uczero |^2
-\half |\nabla_\bfx  \phiczero|^2\Big]d\bfxtld d\bfx \nonumber\\
&+&\int_\Yp  \inttbar_{\bfFhat(\bfx) U_0}
\Big[\Big(\half f^{''}(u_p(\bfFhat(\bfx),\bfxtld))+\phi_p(\bfFhat(\bfx),\bfxtld)\Big)|\uczero |^2\\
&&\hspace{2cm}+ 2u_p(\bfFhat(\bfx),\bfxtld)\uczero  \phiczero+b_c\phiczero \Big]d\bfxtld d\bfx\nonumber \\
&+& \int_\Yp \inttbar_{\bfFhat(\bfx) U_0}\Big[ \frac{\lambda}{2}
(2\nabla_\bfx  \uczero +\nabla_\bfxtld \ucone ) \cdot \nabla_\bfxtld
\ucone -(2\nabla_\bfx  \phiczero+\nabla_\bfxtld \phicone )\cdot
\nabla_\bfxtld \phicone )\Big] d\bfxtld d\bfx. \nonumber \eeqs Since
$\ucone (\bfx,\bfxtld)$ and $\phicone (\bfx,\bfxtld)$ are periodic
on $\bfFhat(\bfx)U_0$ for every $\bfx$, from
equation~\eqref{eq:uphix} we have \beqs\label{eq:uu1phiphi1}
\int_{\bfFhat(\bfx)U_0} \nabla_\bfx  \uczero \cdot \nabla_\bfxtld
\ucone d\bfxtld=0 \qquad \aand \qquad \int_{\bfFhat(\bfx)U_0}
\nabla_\bfx  \phiczero \cdot \nabla_\bfxtld \phicone
d\bfxtld=0\qquad \forall\,\bfx\in \Yp. \eeqs From the min-max
problem~\eqref{eq:minmaxc}, we maximize the expression in
equation~\eqref{eq:E2eta0order} over admissible $\phicone$ and
minimize it over admissible $\ucone$, and obtain \beqs
\label{eq:uphionextld} \nabla_\bfxtld  \ucone (\bfx,\bfxtld)=0\qquad
\aand\qquad \nabla_\bfxtld  \phicone (\bfx,\bfxtld)=0. \eeqs Thus,
it follows that the corrector fields do not exhibit atomic-scale
oscillations up to the second order terms in the multiple scale
expansion~\eqref{eq:phiueta}. Further, from
equation~\eqref{eq:uphionextld}, the last term on the right hand
side of equation~\eqref{eq:E2eta0order} vanishes, and by
equations~\eqref{eq:alphabeta} and \eqref{eq:uphix} we identify the
first two terms in equation~\eqref{eq:E2eta0order} as \beqs
\label{eq:E20}
 \int_{\Yp } \Big[\frac{\lambda}{2} |\nabla_\bfx  \uczero |^2
-\half |\nabla_\bfx  \phiczero|^2 +\gamma(\bfFhat(\bfx))|\uczero |^2
+ 2\alpha(\bfFhat(\bfx))\uczero  \phiczero+b_c\phiczero \Big]d\bfx
=:E_2^{0}(\uczero,\phiczero,\bfy; b_c), \eeqs where the $\gamma$ and
$\alpha$ are defined in equation~\eqref{eq:alphabeta}. In
conclusion,  from the min-max problem in
equation~\eqref{eq:minmaxc}, assuming the multiple scale expansion
given by equation~\eqref{eq:phiueta}, and keeping only the leading
order terms, we have \beqs \label{eq:minmax0}
\calE_2(\bfy;b_c)\approx \min_{\uczero }\max_{\phiczero
}E_2^{0}(\uczero,\phiczero, \bfy; b_c) \eeqs subject to \beqs
\label{eq:calAtld1} \phiczero\in H^1_{per}(\Yp), \quad \uczero \in
H_{per}^1(\Yp),\quad \int_\Yp  (2\alpha(\bfFhat(\bfx))
u_c^0+b_c)d\bfx=0, \eeqs where the constraint on $u_c^0$ follows
from equations~\eqref{eq:calAtld}, \eqref{eq:alphabeta},  and
neglecting higher order terms in equation~\eqref{eq:phiueta}.
Equations \eqref{eq:minmax0}-\eqref{eq:calAtld1} constitute the
governing equations for the corrector fields in their leading order.

We now proceed to derive the governing equations that describe the
elastic response of the defect. From equations~\eqref{eq:Eeps} and \eqref{eq:minmax0} we see that
$\calE^\eps:\calY\to \rz$ defined in equation~\eqref{eq:minmaxeps}
is given by \beqs \label{eq:calEeps1} \calE^\eps( \bfy;
b_c)&=&E(u_\bfy,\phi_\bfy,b_\bfy) +\eps \int_\Yp  b_c\phi_\bfy d\bfx
+\eps^2 \calE_2(\bfy;b_c)+o(\eps^2) \nonumber \\
&=&\int_\Yp  W(\bfFhat(\bfx))d\bfx+\eps \int_\Yp  b_c\phi_\bfy d\bfx+\eps^2
\calE_2(\bfy;b_c)+o(\eps^2 ),
 \eeqs
where in the second equality we have used equation~\eqref{eq:calEW} for $E(u_\bfy,\phi_\bfy,b_\bfy)$. Let
$\bfy^\ast:Y_0\to \Yp$ be the unperturbed minimizer of the outer
minimization problem in \eqref{eq:calE},
\beas
\bfF^\ast(\bfx_0)=\nabla_{\bfx_0}
\bfy^\ast(\bfx_0)\quad \forall\,\bfx_0\in Y_0,\qquad
\bfFhat^\ast(\bfx)=\bfF^\ast( \bfy^{\ast-1}(\bfx)) \quad
\forall\,\bfx\in \Yp, \eeas and, parallel to
equation~\eqref{eq:uphieps}, let \beqs \label{eq:bfyeps}
\bfy^\eps=\bfy^\ast+\eps \bfy_c \eeqs
 with $\bfy^{\eps}\in \calY$ being the minimizer of the outer minimization problem in~\eqref{eq:minmaxeps}. As we are interested in the elastic fields created in response to the perturbation $b_c$, and not the configurational force on $b_c$, we hold the pull back of $b_c$ on to the reference configuration fixed. To this end, we define $\tilde{b}_c(\bfx_0)=b_c(\bfy^{\ast}(\bfx_0))\,\,\forall\, \bfx_0\in Y_0$ as the pull back before introducing the perturbation in the deformation field. Subsequently, for any infinitesimal perturbation of the deformation given by equation~\eqref{eq:bfyeps}, $b_c$ in the current configuration is given by $b_c(\bfx)=\tilde{b}_c(\bfy^{\eps^{-1}}(\bfx)) \,\,\forall\, \bfx\in \Yp$.
Further, let \beas
\bfF^\eps(\bfx_0)=\nabla_{\bfx_0}
\bfy^\eps(\bfx_0)\quad \forall\,\bfx_0\in Y_0,\qquad
\bfFhat^\eps(\bfx)=\bfF^\eps( \bfy^{\eps^{-1}}(\bfx)) \quad
\forall\,\bfx\in \Yp, \eeas
 and
\beqs \label{eq:bfCbfB}
[\bfC(\nabla_{\bfx_0}\bfy^\ast)]_{piqj}:=\frac{\partial^2 J
W}{\partial \bfF_{pi}\partial
\bfF_{qj}}(\nabla_{\bfx_0}\bfy^\ast),\qquad
[\bfB(\nabla_{\bfx_0}\bfy^\ast)]_{pi}:=\frac{\partial
J\beta}{\partial \bfF_{pi}}(\nabla_{\bfx_0}\bfy^\ast). \eeqs
Since
\beas
\int_Y  W(\bfFhat^\eps(\bfx))d\bfx=\int_{Y_0}  J(\bfF^\eps(\bfx_0))W(\bfF^\eps(\bfx_0))d\bfx_0,
\eeas
we have
\beqs \label{eq:WFhatexpand}
\int_Y  W(\bfFhat^\eps(\bfx))d\bfx 
=\int_{Y_0}
J(\nabla_{\bfx_0}\bfy^\ast)W(\nabla_{\bfx_0}\bfy^\ast)d\bfx_0 +\eps
\int_{Y_0}\nabla_{\bfx_0} \bfy_c \cdot \bfS(\nabla_{\bfx_0}\bfy^\ast) d\bfx_0 \nonumber \\
+\eps^2 \int_{Y_0}  \half \nabla_{\bfx_0}\bfy_c \cdot
\bfC(\nabla_{\bfx_0}\bfy^\ast) \cdot \nabla_{\bfx_0}\bfy_c d\bfx_0+o(\eps^2).
\eeqs
Further, by equations \eqref{eq:f2sint} and \eqref{eq:alphabeta} we have
\beqs \label{eq:bcphiexpand}
 \int_\Yp  b_c\phi_{\bfy^\eps} d\bfx &\approx& \int_\Yp b_c(\bfx) \inttbar_{\bfFhat^\eps(\bfx)U_0} \phi_p(\bfFhat^\eps(\bfx), \bfxtld) d\bfxtld d\bfx =
 \int_{Y_0} \tilde{b}_c(\bfx_0) J (\bfF^\eps) \beta (\bfF^\eps) d \bfx_0 \nonumber \\
 &=& \int_{Y_0}\tilde{b}_c(\bfx_0) J (\nabla \bfy^\ast) \beta (\nabla \bfy^\ast) d \bfx_0+\eps\int_{Y_0}\tilde{b}_c(\bfx_0) \nabla_{\bfx_0} \bfy_c \cdot \bfB(\nabla \bfy^\ast) d \bfx_0+o(\eps)\nonumber\\
 &\approx& \int_Y b_c\phi_{\bfy^\ast} d\bfx + \eps\int_{Y_0}\tilde{b}_c \nabla_{\bfx_0} \bfy_c \cdot \bfB(\nabla \bfy^\ast) d \bfx_0.
\eeqs
Replacing $\bfy$ in equation~\eqref{eq:calEeps1} by
$\bfy^\eps$ given by equation~\eqref{eq:bfyeps}, expanding and
keeping terms up to $\eps^2$, by equations \eqref{eq:WFhatexpand}
and \eqref{eq:bcphiexpand} we obtain \beqs \label{eq:calEeps2}
\calE^\eps(\bfy^\ast+\eps \bfy_c; b_c) &\approx&\int_{Y_0}
J(\nabla_{\bfx_0}\bfy^\ast)W(\nabla_{\bfx_0}\bfy^\ast)d\bfx_0 +\eps
\int_{Y_0}\nabla_{\bfx_0} \bfy_c \cdot
\bfS(\nabla_{\bfx_0}\bfy^\ast) d\bfx_0 +\eps
\int_Yb_c\phi_{\bfy^\ast} d\bfx
 \nonumber \\
&&+\eps^2 \int_{Y_0} \Big\{ \half \nabla_{\bfx_0}\bfy_c \cdot
\bfC(\nabla_{\bfx_0}\bfy^\ast) \cdot \nabla_{\bfx_0}\bfy_c
+\tilde{b}_c \nabla_{\bfx_0}\bfy_c \cdot \bfB(\nabla_{\bfx_0}\bfy^\ast)\Big\}d\bfx_0\nonumber\\
&&+\eps^2 \calE_2(\bfy^\ast;b_c)+o(\eps^2), \eeqs
Since
$\bfy^\ast$ is a local minimizer satisfying equation~\eqref{eq:ELelastic}, it follows that $\int_{Y_0}\nabla_{\bfx_0}\bfy_c
\cdot \bfS(\nabla_{\bfx_0}\bfy^\ast)d\bfx_0=0$. We further neglect
$o(\eps^2)$-term in equation~\eqref{eq:calEeps2}. Finally, the outer
minimization problem given by equation~\eqref{eq:minmaxeps} reduces
to a minimization problem on $\bfy_c$ and is given by \beqs
\label{eq:minyc} \calE_{el}(b_c):=\min_{\bfy_c\in \calY} \int_{Y_0}
\Big\{ \half \nabla_{\bfx_0}\bfy_c \cdot \bfC(\nabla_{\bfx_0}\bfy^\ast)
\nabla_{\bfx_0}\bfy_c +\tilde{b}_c \nabla_{\bfx_0}\bfy_c \cdot
\bfB(\nabla_{\bfx_0}\bfy^\ast) \Big\}d\bfx_0. \eeqs Thus, a minimizer $\bfy_c$
satisfies the following Euler-Lagrange equation which constitutes
the governing equation for the elastic response in the presence of a
defect \beqs \label{eq:elastic} \Diverg
[\bfC(\nabla_{\bfx_0}\bfy^\ast)
\nabla_{\bfx_0}\bfy_c+\tilde{b}_c
\bfB(\nabla_{\bfx_0}\bfy^\ast)]=0\qquad \oon\;Y_0. \eeqs

An important quantity in the study of defects is the defect
formation energy, which is defined as the excess energy  in the
system with a defect measured from a reference state of a perfect
crystal consisting of same number of particles---in this case the
number of electrons and nuclei. In the framework of the present study,
it is given by \beas
\calE_{d}(b_c):=[\calE_{tot}(b_c)-\calE_{tot}(0)-\eps\int_Y
b_c\phi_{\bfy^\ast}d\bfx]/\eps^2. \eeas From the previous discussions, the defect formation energy (defect energy) can be expressed, to the leading order, as
the following min-min-max (saddle point) problem: \beqs
\label{eq:calEv} \calE_{d}(b_c)\approx \min_{\bfy_c} \min_{u_c^0}
\max_{\phi_c^0  } \int_{Y_0} \Big \{ \half \nabla_{\bfx_0} \bfy_c
\cdot \bfC(\nabla_{\bfx_0} \bfy^\ast)
 \nabla_{\bfx_0} \bfy_c+\tilde{b}_c \nabla_{\bfx_0} \bfy_c \cdot \bfB(\nabla_{\bfx_0} \bfy^\ast) \Big \}d\bfx_0\nonumber \\
+\int_\Yp \Big \{\gamma(\bfFhat^\ast)|\uczero |^2 +\frac{\lambda}{2}
|\nabla_\bfx  \uczero |^2 -\frac{1}{2 } |\nabla_\bfx  \phiczero|^2+2
\alpha( \bfFhat^\ast)\uczero  \phiczero+b_c\phiczero \Big \}d\bfx
\eeqs subject to \beqs \label{eq:calAtld2} \phiczero\in
H^1_{per}(\Yp), \quad \uczero \in  H_{per}^1(\Yp),\quad \int_\Yp
(2\alpha(\bfFhat^\ast) u_c^0+b_c)d\bfx=0, \quad\bfy_c\in \calY.
\eeqs Associated with the above min-min-max problem, the
Euler-Lagrange equations are the elasticity
equation~\eqref{eq:elastic} for $\bfy_c$  on the reference
configuration $Y_0$ and \beqs \label{eq:ELall} \becs
\Delta  \phiczero+  2 \alpha(\bfFhat^\ast) \uczero +b_c=0&\oon\;\Yp,\\
-\lambda \Delta \uczero + 2 \gamma(\bfFhat^\ast)\uczero +2
\alpha(\bfFhat^\ast)
\phiczero=0 &\oon\;\Yp\\
\eecs
 \eeqs
 for  $(\phiczero, \uczero)$ on the current configuration $\Yp$.
Note that the elasticity problem \eqref{eq:elastic} for $\bfy_c$ is
not coupled with the equations for $(\phiczero, \uczero)$. In terms
of the solutions  $(\bfy_c, \phiczero, \uczero)$ to equations
\eqref{eq:elastic} and \eqref{eq:ELall}, the defect energy can be
written as \beqs \label{eq:calEd1} \calE_d(b_c)\approx\half \int_{Y_0}
\tilde{b}_c \nabla \bfy_c \cdot \bfB(\nabla \bfy^\ast) + \half
\int_Y b_c \phi_c^0. \eeqs

\section{Far fields}\label{sec:farFields}

In this section we determine the far-field behavior of the fields
$(\phiczero,\uczero,\bfy_c)$ from the governing  equations in
\eqref{eq:ELall} that will aid in determining the optimal
coarse-graining rates for these fields. In this analysis, we assume
$b_c$ is continuous, bounded and
 supported within the ball $B_{r_0}= \{\bfx:|\bfx|\le {r_0}\}$. Although the analysis in the previous
section was performed on the supercells $Y_0$ and $\Yp$, we note
that the results of the analysis are independent of the supercells
and thus to determine the asymptotic behavior of the corrector
fields we assume $Y_0=\Yp=\rz^3$. We first calculate the far field
behavior of $(\phiczero,\uczero)$ for a homogeneous deformation with
$\nabla_{\bfx_0}\bfy^\ast=\bfF_0\in \rz^{3\times 3}$ on $\rz^3$. In
this case, $\bfFhat^\ast=\bfF_0$ on $\rz^3$ as well;
$\bfC(\nabla_{\bfx_0}\bfy^\ast) $, $
\bfB(\nabla_{\bfx_0}\bfy^\ast)$,  $\alpha(\bfFhat^\ast)$,
$\gamma(\bfFhat^\ast)$ are constants on $\rz^3$ and we drop their
dependence on $\bfF_0$ in notation.
 Further, the periodic boundary conditions in equation
\eqref{eq:calAtld2} shall be replaced by appropriate decay
conditions at the infinity. Dropping the subscript $_c$ and
superscript $^0$ in $(\phiczero,\uczero, \bfy^c)$ in equations
\eqref{eq:ELall} and \eqref{eq:elastic}, we rewrite our problem
for ($\phi, u, \bfy$) as \beqs \label{eq:ELall1} \becs
\Delta  \phi+  2 \alpha u +b_c=0&\oon\;\rz^3,\\
-\lambda \Delta u + 2 \gamma u +2 \alpha \phi=0 &\oon\;\rz^3,\\
\Diverg [\bfC \nabla_{\bfx_0}\bfy+b_c(\bfF_0\bfx_0) \bfB]=0 &\oon\;\rz^3,\\
\eecs
 \eeqs
 subject to
 \beqs \label{eq:uphiinfitycond}
|\phi(\bfx)|,\; |u(\bfx)|,\;| \bfy|\to 0\quad\aas\; |\bfx|\to
+\infty, \qquad \int_{\rz^3}  (2\alpha u+b_c)=0. \eeqs

We now address the solutions of the first two of equation \eqref{eq:ELall1}.
 Taking Laplacian of equation $\eqref{eq:ELall1}_2$
and inserting into equation $\eqref{eq:ELall1}_1$, we obtain \beqs
\label{eq:pdeu}
 \Delta \Delta u-\frac{2}{l_1^2} \Delta u+\frac{1}{l_0^4} u=b
\qquad \oon\;\rz^3, \eeqs
where $l_0>0$, $\Real(l_1)\ge 0$,
\beqs \label{eq:c1c0}
\frac{1}{l_1^2}=\frac{\gamma}{\lambda}, \quad \frac{1}{l_0^4}=\frac{4\alpha^2}{\lambda}>0, \quad b=
-\frac{2\alpha b_c}{\lambda}. \eeqs

The constants $l_1, \, l_0$ determine the asymptotic behavior of the
fundamental solution at the infinity.
Since equation~\eqref{eq:pdeu} is
linear, we express its solution as \beqs \label{eq:pdeusol}
u(\bfx)=(E_u \ast b)(\bfx)=\int_{\rz^3} E_u(\bfx-\bfx') b(\bfx')d\bfx',
\eeqs where $E_u$ is the fundamental solution satisfying
 \beas (\Delta \Delta -\frac{2}{l_1^2} \Delta +\frac{1}{l_0^4} )E_u=\delta(0),
 \eeas
 and $\delta(0)$ is the Dirac distribution.
We find the fundamental solution $E_u$ by Fourier analysis. Solving the
algebraic equation \beqs \label{eq:xeq} x^4+\frac{2}{l_1^2}x^2+\frac{1}{l_0^4}=0, \eeqs
we obtain two roots $\kappa_\pm$ with $\Imag(\kappa_\pm)\ge 0$ and satisfying
\beqs \label{eq:kappa}
\kappa_\pm^2=-\frac{1}{l_1^2}\pm \sqrt{\frac{1}{l_1^4}-\frac{1}{l_0^4}}\;. \eeqs
The two other roots with $\Imag(\kappa_\pm)< 0$ are discarded as they will correspond to exponentially growing
solutions in $(u,\phi)$, defined subsequently, and do not satisfy the decay conditions imposed in \eqref{eq:uphiinfitycond}.
By Fourier analysis, we
have \beqs \label{eq:Fsol0}
E_u(\bfx)&=&\frac{1}{(2\pi)^3}\int_{\rz^3}\frac{1}{|\bfk|^4+\frac{2}{l_1^2}|\bfk|^2+\frac{1}{l_0^4}}\exp(i\bfk\cdot \bfx)d\bfk \nonumber \\
&=&\frac{1}{(2\pi)^3}\int_{\rz^3}\frac{1}{(\kappa_+^2-\kappa_-^2)}\Big[\frac{1}{|\bfk|^2-\kappa_+^2}-
\frac{1}{|\bfk|^2-\kappa_-^2}\Big]\exp(i\bfk\cdot \bfx)d\bfk. \eeqs
We are therefore motivated to consider the fundament solution of the
operator \beqs \label{eq:helmoltz} (\Delta +\kappa^2)G =\delta(0)
\eeqs for some $\kappa\in \cz$ with $\Imag(\kappa)\ge 0$. By the
standard method (cf. \cite{Jackson1999} page 243), we have \beqs
\label{eq:Gs}
 G(\bfx, \kappa)=
 \becs
 -\frac{\exp( i\kappa |\bfx|)}{ 4\pi|\bfx|} &\iif\;\Imag(\kappa)\neq 0,\\
 -A\frac{\exp( i\kappa |\bfx|)}{ 4\pi|\bfx|}-B\frac{\exp(-i\kappa |\bfx|)}{ 4\pi|\bfx|} &\iif\;\Imag(\kappa)=0,
 \eecs
\eeqs where $A+B=1$.
 In Fourier space,  equation~\eqref{eq:helmoltz} can be rewritten as
\beqs \label{eq:GsF}
G(\bfx,\kappa)=\frac{1}{(2\pi)^3}\int_{\rz^3}\frac{-1}{|\bfk|^2-\kappa^2}\exp(i\bfk\cdot
\bfx)d\bfk. \eeqs
If $|l_1|\neq l_0$, i.e., $\kappa_+^2\neq\kappa_-^2$,
from equations \eqref{eq:Fsol0}-\eqref{eq:GsF} we have \beqs \label{eq:Fsol1}
E_u(\bfx)=\frac{-1}{(\kappa_+^2-\kappa_-^2)}\Big[ G(\bfx,
\kappa_+)-G(\bfx, \kappa_-)\Big].
 \eeqs
 If $|l_1|=l_0>0$, i.e.,
$\kappa_+^2=\kappa_-^2=-\frac{1}{l_1^2}$, sending $\kappa_+$ to
$\kappa_-$ in equation \eqref{eq:Fsol1} we obtain
 \beqs \label{eq:Fsol2}
  E_u(\bfx)= \becs
\frac{-1}{2\kappa}\partial_\kappa G(\bfx, \kappa)=
\frac{l_0}{8\pi  } \exp(-|\bfx|/l_0)&\iif\;\gamma>0,\;\kappa=i/l_0,\\
\frac{-1}{2\kappa}\partial_\kappa G(\bfx, \kappa)= \frac{i A l_0}{8\pi}
 \exp(i|\bfx|/l_0)+\frac{iB l_0}{8\pi } \exp(-i|\bfx|/l_0)
&\iif\;\gamma<0,\;\kappa=1/l_0.\\
\eecs \eeqs
Further, by the second of equation \eqref{eq:ELall1},  the associated potential is
given by \beqs \label{eq:phi}
\phi(\bfx)=\frac{\lambda}{2\alpha} \Delta u-\frac{\gamma}{\alpha} u=\int_{\rz^3} E_{\phi}(\bfx-\bfx') b(\bfx')d\bfx',
\eeqs
where
\beqs \label{eq:Ephi}
E_\phi(\bfx)=\frac{\lambda}{2\alpha}[ \Delta E_u(\bfx) -\frac{2}{l_1^2} E_u(\bfx)].
\eeqs
We remark that the above formal calculations can be
rigorously justified (cf. e.~g. \cite{Rudin1991}, chapter 7).

Note that the last of equation \eqref{eq:uphiinfitycond} requires
\beqs
\label{eq:uconst} \int_{\rz^3}  (\frac{1}{l_0^4} u-b)=0.
 \eeqs
If $\int b(\bfx) d\bfx\neq 0$ and $b(\bfx)$ has a compact support in a ball around the origin, then for large $|\bfx|$
the solution $u(\bfx)$ is well approximated by the Green's function $E_u$ in equation~\eqref{eq:Fsol1}, which is not integrable if $\kappa_\pm$ are real numbers as $\int_{\rz^3}\frac{\exp(ik|\bfx|)}{|\bfx|}d\bfx$ is not integrable for real $k$.
We therefore conclude that $\kappa_\pm$ should  be  both
nonreal numbers. This is possible for the following three cases:
\begin{enumerate}
  \item  $\gamma>0$ and $l_0>l_1$. In this case, all roots of equation \eqref{eq:xeq} are pure imaginary.
   By equations \eqref{eq:Fsol1} and~\eqref{eq:Ephi}, we have
\beqs \label{eq:u1} 
&&E_u(\bfx)=\frac{1}{4\pi
(\kappa_+^2-\kappa_-^2)|\bfx|}\Big[ \exp(i \kappa_+
|\bfx|)-\exp(i\kappa_-|\bfx|)\Big], \nonumber \\
 &&E_\phi(\bfx)=\frac{1}{4\pi
(\kappa_+^2-\kappa_-^2)|\bfx|}\Big[ C_+\exp(i \kappa_+
|\bfx|)-C_-\exp(i\kappa_-|\bfx|)\Big],
 \eeqs
 where
 \beqs \label{eq:Cpm}
C_\pm=\frac{\lambda}{2\alpha} (-\kappa_\pm^2-\frac{2}{l_1^2})=
\frac{\lambda}{2\alpha}\kappa_\mp^2.
\eeqs

\item   $\gamma>0$ and $l_0=l_1$. This is the first case in equation \eqref{eq:Fsol2} and we have
\beqs \label{eq:u2} 
E_u(\bfx)=\frac{l_0}{8\pi  } \exp(-|\bfx|/l_0),\qquad
E_\phi(\bfx)=\frac{-\lambda}{16 \pi \alpha  }(\frac{1}{l_0}+\frac{2}{r}) \exp(-|\bfx|/l_0).
 \eeqs
  \item  $l_0<|l_1|$. In this case, all roots of equation \eqref{eq:xeq} are nonreal
  and the fundamental solutions are given by \eqref{eq:u1} as well.

\end{enumerate}
To verify the constraint~\eqref{eq:uconst}, we integrate equation
\eqref{eq:pdeu} on the ball $B_N$ with radius $N$, and by the
divergence theorem arrive at \beqs \label{eq:u4} \int_{\partial
B_N}(\nabla \Delta u-\frac{2}{l_1^2}\nabla u) \cdot \bfxhat
dS+\int_{B_N} (\frac{1}{l_0^4}u-b)d\bfx=0, \eeqs where
$\bfxhat=\bfx/|\bfx|$. Sending $N\to +\infty$, we arrive at equation
\eqref{eq:uconst} since the first term in the above equation
vanishes for expressions in equations \eqref{eq:u1} or \eqref{eq:u2}.

Finally, we remark that the solution to the last of
equation \eqref{eq:ELall1} is given by the classic theory of linear
elasticity (cf. e.~g.~\cite{Mura1987}, chapter 1).

Equations~\eqref{eq:pdeusol}, \eqref{eq:phi}, \eqref{eq:u1}-\eqref{eq:u2} and the theory of elasticity determine
the far-field behavior of $(u,\phi, \bfy)$, where the perturbation $b_c$
plays the role of a source. For a continuous bounded $b_c$ supported within the ball $B_{r_0}=\{\bfx:|\bfx|<r_0\}$, we
have that for some $R>0$ and some $C,\kappa>0$,
 \beqs \label{eq:farfields}
|u(\bfx)|\le C\exp(-\kappa |\bfx|), \quad |\phi(\bfx)|\le
 C\exp(-\kappa |\bfx|), \quad |\bfy(\bfx_0)|\le \frac{C}{|\bfx|^2}\qquad
\forall\,|\bfx|>R.
 \eeqs
With the above estimates on the far-fields, we continue our
solutions to equations \eqref{eq:ELall1}-\eqref{eq:uphiinfitycond} for a
particular example in the next section.

\section{Defect energy and cell-size effects}\label{sec:defectCell-size}
In this section we study how the defect energy
depends on the size of the supercell.
For simplicity, we assume that the supercell is the  ball $B_{R_0}=\{\bfx:|\bfx|<R_0\}$,
the coefficients $l_0>l_1>0$ and thus both the roots
$\kappa_\pm$ in equation \eqref{eq:kappa} are pure imaginary
numbers. We denote by \beqs \label{eq:kpm}
k_\pm=-i\kappa_\pm=\sqrt{\frac{1}{l_1^2}\mp
\sqrt{\frac{1}{l_1^4}-\frac{1}{l_0^4}}} =\sqrt{\frac{\gamma
\mp\sqrt{\gamma^2-4\alpha^2 \lambda}}{\lambda}}>0.
 \eeqs
Below we solve equations \eqref{eq:ELall1} for the corrector fields $(u, \phi, \bfy)$ with
 \beas b(\bfx)=\becs
\rho &\iif\;|\bfx|\le {r_0},\\
0&\iif\;|\bfx|> {r_0}, \eecs
\eeas where $\rho \in \rz$ is a constant, $r_0<R_0$ describes the
length scale of the defect representative of a vacancy.
We apply the Dirichlet boundary condition
\beqs \label{eq:ubcR0} u(\bfx)=0, \quad \phi(\bfx)=\frac{\lambda}{2\alpha} \Delta u(\bfx)-\frac{\gamma}{\alpha} u(\bfx)=\varsigma, \quad \bfy=0\qquad \oon\;\partial B_{R_0},
\eeqs
where $\varsigma\in \rz$ is a constant determined by
the constraint~\eqref{eq:uconst}.

We first consider the electrostatic contribution of the defect energy, i.e.,
the second term on the r.h.s. of equation~\eqref{eq:calEd1}.
By symmetry, we have
$u=u(r)$ with $r=|\bfx|$. Therefore, equation~\eqref{eq:pdeu}
can be rewritten as \beas \frac{d^4}{dr^4}ru-\frac{2}{l_1^2}
\frac{d^2}{dr^2}ru+\frac{1}{l_0^4} ru=rb \qquad \forall\,0<|\bfx|<R_0. \eeas
From the theory of ordinary differential equation, we obtain \beas
ru(r)= \becs {\rho r l_0^4}  +
C_1\exp(\kk_+ r)+C_2\exp(\kk_- r)\\
\hspace{3cm}+C_3\exp(-\kk_+ r)+C_4\exp(-\kk_-r)&\iif\;r\le {r_0},\\
C_5\exp( \kk_+ r)+C_6\exp(\kk_- r) \\
\hspace{3cm}+C_7\exp(-\kk_+ r)+C_8\exp(-\kk_-r)&\iif\;r\ge {r_0},\\
\eecs \eeas where the constants $C_i$ ($i=1,\cdots, 8$) are
determined by the analyticity of $u(\bfx)$ at $r=0$ (which implies
$u(r)$ is an even function, i.e., $C_1+C_3=0$ and $C_2+C_4=0$), the
continuities of $\frac{d^m}{dr^m}(ru)$ for $m=0,1,2,3$ at $r={r_0}$,
the boundary condition~\eqref{eq:ubcR0}  and the
constraint~\eqref{eq:uconst}. Direct calculations reveal that  these
conditions imply
 \beqs \label{eq:C1C2}
\begin{bmatrix}
1 \quad 0\quad 1\quad 0 &0\quad 0 \quad 0 \quad 0\\
0 \quad 1\quad 0\quad 1 &0\quad 0 \quad 0 \quad 0\\
\bfa(r_0,0) &-\bfa(r_0,0)\\
\bfa(r_0,1) &-\bfa(r_0,1)\\
\bfa(r_0,2) &-\bfa(r_0,2)\\
\bfa(r_0,3) &-\bfa(r_0,3)\\
0 \quad 0\quad 0\quad 0 &\bfa(R_0,0)\\
0 \quad 0\quad 0\quad 0 &
\begin{array}{c}
R_0\bfa(R_0,3)-\bfa(R_0,2)\\
-\frac{2}{l_1^2}R_0 \bfa(R_0,1)+\frac{2}{l_1^2} \bfa(R_0,0)\\
\end{array}
\end{bmatrix}
\begin{bmatrix}
C_1\\
C_2\\
C_3\\
C_4\\
C_5\\
C_6\\
C_7\\
C_8\\
\end{bmatrix}=
\begin{bmatrix}
0\\
0\\
-\rho r_0 l_0^4\\
-\rho l_0^4 \\
 0\\
0\\
0\\
0\\
\end{bmatrix},
\eeqs
where the $1\times 4$ row vector $\bfa(r,m)$ is given by
\beas
&&\bfa(r,m)=[k_+^m \exp(k_+ r),\; k_-^m \exp(k_- r),\; (-k_+)^m \exp(-k_+ r), \;(-k_-)^m \exp(-k_- r)].
\eeas
Note that the last row of equation \eqref{eq:C1C2} follows from setting the ball $B_N$
to be $B_{R_0}$ in equation~\eqref{eq:u4} and  the constraint~\eqref{eq:uconst}.
 Further, from equation \eqref{eq:phi} we have $ r\phi(r)=\frac{\lambda}{2\alpha}  \frac{d^2}{dr^2} ru-\frac{\gamma}{\alpha} r u$
 and hence
 \beqs \label{eq:rphi}
 r\phi(r)=
\becs
- \frac{\rho \gamma \lambda }{4\alpha^3  }r
+ C_+C_1[\exp(\kk_+ r)-\exp(-\kk_+ r)]\\
\hspace{2cm}+C_-C_2[\exp(\kk_- r)-\exp(-\kk_-r)] &\iif\;r\le {r_0},\\
C_+C_5\exp(\kk_+ r)+C_-C_6\exp(\kk_- r)\\
\hspace{2cm}+C_+C_7\exp(-\kk_+ r)+C_-C_8\exp(-\kk_- r)&\iif\;r> {r_0},\\
\eecs
\eeqs
where, by equations \eqref{eq:Cpm} and \eqref{eq:kpm}, $C_\pm=-\lambda k_\mp^2/2\alpha$.
Therefore, the electrostatic contribution to  the defect energy
is given by
\beqs \label{eq:calEden}
\calE_d^{es}(R_0):=\half \int_{\rz^3} (-\frac{\lambda}{2\alpha} b) \phi =\frac{-\lambda}{4\alpha}\Big\{\frac{-\pi \gamma \lambda  \rho r_0^3 }{3\alpha^3  }
+C_+C_1\frac{8\pi [k_+r_0 \cosh(k_+r_0)-\sinh(k_+r_0) ]}{k_+^2} \nonumber\\
+C_-C_2\frac{8\pi [k_-r_0 \cosh(k_-r_0)-\sinh(k_-r_0) ]}{k_-^2}
\Big\}.
\eeqs

We remark that the algebraic equations \eqref{eq:C1C2}
determine the constants $[C_1,\cdots, C_8]$ uniquely.
Analytical expressions of these constants are desirable but impractical to write them down. In the
limit $R_0\to +\infty$, we find
 \beqs \label{eq:C1C2sol}
&& C_1=\frac{ \rho l_0^4 k_-^2(1+k_+r_0)\exp(-k_+r_0)}{2\kk_+ (k_+^2-k_-^2)},
 \qquad  C_2=\frac{- \rho l_0^4 k_+^2(1+k_-r_0)\exp(-k_-r_0)}{2\kk_- (k_+^2-k_-^2)} \nonumber \\
&& C_5=C_6=0,\qquad C_7=\frac{\rho l_0^4 k_-^2[k_+r_0\cosh(k_+r_0)-\sinh(k_+r_0)]}{  \kk_+ (k_+^2-k_-^2)},\\
 && C_8=\frac{\rho l_0^4 k_+^2[k_-r_0\cosh(k_-r_0)-\sinh(k_-r_0)]}{ \kk_- (k_+^2-k_-^2)}. \nonumber
 \eeqs

\begin{figure}[htbp]
    \begin{center}
      {\scalebox{0.8}{\includegraphics{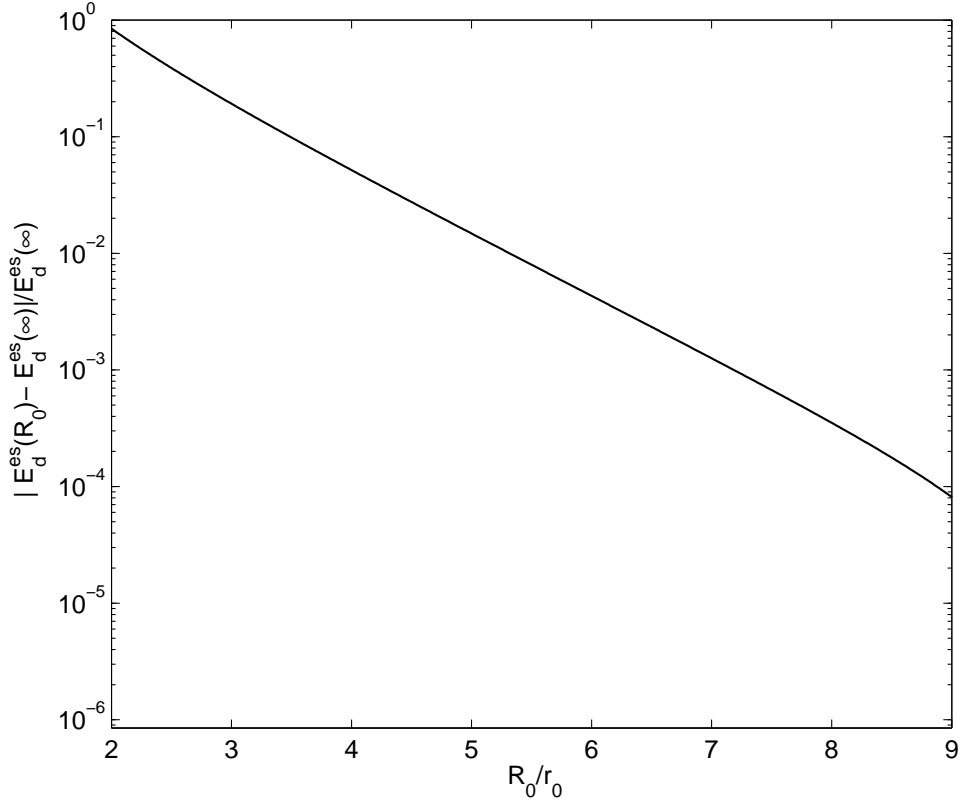}}}
       \caption{Cell-size effects showing relative error in the electrostatic contribution to the defect energy from finite cell calculations.\bigskip}
      \label{fig:Cell-Size}
      \end{center}
\end{figure}

For general cases with finite $R_0$, which represent computations on
a finite simulation cell,  we resort to numerical solutions. In
particular, we are interested in estimating the error incurred in
the defect energy from using a simulation cell, and its
dependence on the cell-size. To this end, we have conducted a
periodic calculation on a unit cell of FCC lattice for aluminum using a real-space formulation for
OFDFT and a finite-element discretization of the formulation \citep{GKBO2007}. In
our simulation, we used the TFW family of kinetic energy functionals
with $\lambda=\frac{1}{6}$ and a modified form of Heine-Abarenkov
pseudopotential for aluminium \citep{Goodwin}. We subsequently
estimate the constants $\alpha,\beta,\gamma$ from our numerical calculations to be \beas \alpha =
0.1629,\quad \beta = -0.0509,\quad \gamma=0.9449. \eeas We now
estimate the cell-size effects in the electrostatic contribution to
the energy of a defect that is representative of a vacancy.
A reasonable choice for the length scale of a
vacancy is $r_0=a_0/2$, where $a_0$ is the lattice parameter for
aluminum which is computed to be 7.5 a.u. Using
equations~\eqref{eq:C1C2}-\eqref{eq:calEden}, we numerically solve
for the electrostatic contribution to defect energy. Figure~\ref{fig:Cell-Size} shows
our estimate of cell-size effects from finite cell simulations.
As is evident from these results, $R_0=6r_0=3a_0$
is necessary for the approximation errors from finite cell-size
studies to be within $1\%$ of the defect energy---a
threshold representative of chemical accuracy. In typical electronic
structure simulations this $R_0$ corresponds to a simulation cell
with $6\times6\times6$ FCC unit cells containing 864 aluminum atoms.
This estimate is in close agreement with recent cell-size studies on
vacancy formation energies conducted in \citet{GBO2007}, where about
$10^3$ atoms were required for the cell-size effects in defect
formation energy to be within 0.01eV. We note that despite the
exponential decay in the electronic fields, cell-size effects are significant,
even for a simple defect like vacancy. In the more
accurate models of density functional theory, like the Kohn-Sham
formulation, the decay in electronic fields is known to be slower
and hence cell-sizes beyond those considered in previous electronic
structure studies may be needed for an accurate study of defects.

We now consider the elastic contribution of the defect energy, i.e., the first term on the r.h.s. of equation~\eqref{eq:calEd1}, which is a standard calculation and provide it for the sake of completeness. For simplicity, we assume that the stiffness tensor of the crystal, defined by \eqref{eq:bfCbfB}, is isotropic and that the ``eigenstress" $\bfB$ is dilatational.
Let $\mu$ be the shear modulus,  $\kappa$ be the bulk modulus, and $\bfB=\sigma_0\bfI$ ($\bfI$ is the identity matrix).
Based on the Eshelby's solution (Eshelby 1957), we find that the displacement is given by
\beas
 \bfy=\nabla \xi,\qquad \xi=\becs
\half \Theta_1 r^2 +\Theta_0 &\iif\;|\bfx|\le {r_0},\\
\half \Theta_2 r^2+\Theta_3 \frac{1}{r}&\iif\;r_0<|\bfx|< {R_0},
\eecs
\eeas
where $\Theta_1,\Theta_2,\Theta_3\in \rz$ are constants to be determined.
Indeed, by direct calculations we verify that
the function $\bfy$ given by the above expression  satisfies the last of equation \eqref{eq:ELall1}
inside the ball $r<r_0$ and inside the annulus region $r_0<r<R_0$.
Across the interface $r=r_0$, the continuity of $\bfy$,
the continuity of traction and the boundary condition $\bfy=0$ at $r=R_0$
imply
 \beas
\Theta_1=\Theta_2-\Theta_3/r_0^3,\quad \rho \sigma_0 + 3\kappa \Theta_1=3 \kappa \Theta_2+4\mu \Theta_3/r_0^3, \quad
\quad \Theta_2-\Theta_3/R_0^3=0.
 \eeas
 Direct calculation reveals that \beas
 \Theta_1=\frac{\rho \sigma_0}{4\mu +3\kappa} (\frac{r_0^3}{R_0^3}-1).
 \eeas
 Therefore, the elastic contribution to the defect energy is given by
 \beas
\calE_d^{el}(R_0):=\half \int_{r\le r_0} \rho \nabla \bfy \cdot \sigma_0 \bfI=
\frac{3 \rho \sigma_0}{2} \Theta_1=\frac{3\rho^2 \sigma_0^2}{2(4\mu +3\kappa)} (\frac{r_0^3}{R_0^3}-1).
 \eeas
The elastic contribution of the defect energy has a slower asymptotic decay ($O(\frac{1}{R_0^3})$) in comparison to the electronic contribution and is one other reason to consider large cell-sizes to ensure the accurate computation of the energetics of defects.

\section{Extensions}\label{sec:extension}

The form of OFDFT energy we considered for the multiple scale
analysis in prior sections represents an orbital-free model with
TFW kinetic energy functionals without exchange and correlation
terms. In this section we comment on other general forms of energies
that are widely used in OFDFT computations. We remark that the
multiple scale analysis is independent of the form of the non-linear
term $f(u)$ appearing in equation~\eqref{eq:Eb}, and thus including the
exchange and correlation energies does not affect the analysis or the
derived expressions. However, the non-local kernel energies can not be
represented by a local function of the form $f(u)$, and we now present the
extension of our analysis to these commonly used kinetic energy functional forms.

The OFDFT formulations employed in numerical studies widely use functional forms for
kinetic energy that are non-local in real-space, called \emph{kernel
energies}, which are considered to be more accurate than the local
TFW functionals (cf.~equation~\eqref{eq:TFW}). We refer to
\cite{WT1992,SM1994,WGC1998,WGC1999} for further details on these
models. We also remark that recent analysis \citep{BC2005} has shown
that some of the proposed models lack global stability and can pose
serious numerical issues. For the sake of completeness, we briefly
discuss the multiple scale analysis of these non-local kernel
energies. The functional form of these kernel energies is given by
\beqs \label{eq:EKer} E^{Ker}(u)=
\int_{\Yp}\int_{\Yp}p(u(\bfx))K(|\bfx-\bfx'|)q(u(\bfx'))d\bfx
d\bfx', \eeqs where $p(u),q(u)$ are functions whose specific form
depends on the particular flavor of the OFDFT model, and the total energy is given by \beqs \label{eq:Eb1}
E(\phi, u;b_{\bfy})=  \int_{\Yp } \Big[f(u)+\frac{1}{2} |\nabla u|^2
-\frac{1}{2} |\nabla \phi|^2+(u^2+b_{\bfy})\phi
\Big]d\bfx+E^{Ker}(u). \eeqs We define the following potentials
which will be used to reformulate the non-local kernel energy given
by equation~\eqref{eq:EKer} into a local variational problem: \beqs
V_{p}(\bfx)=\int_{\Yp}K(|\bfx-\bfx'|)p(u(\bfx'))d\bfx',\qquad
V_{q}(\bfx)=\int_{\Yp}K(|\bfx-\bfx'|)q(u(\bfx'))d\bfx'. \eeqs Taking
the Fourier transform of the above expressions we obtain
\beqs\label{eq:VpVqFourier}
\hat{V}_{p}(\mathbf{k})=\hat{K}(|\mathbf{k}|)\hat{p}(\mathbf{k}),\qquad
\hat{V}_{q}(\mathbf{k})=\hat{K}(|\mathbf{k}|)\hat{q}(\mathbf{k}).
\eeqs

 Following the ideas developed in \citet{CK2002}, $\hat{K}$ can
be modeled to good accuracy using a sum of partial fractions of the
form, \beqs\label{eq:KerFourier} \hat{K}(|\mathbf{k}|)\approx
\sum_{j=1}^{m}\frac{P_j|\mathbf{k}|^2}{|\mathbf{k}|^2+Q_j} \eeqs
where $P_{j}$, $Q_j$, $j=1\ldots m$ are constants which are fitted
to best reproduce $\hat{K}(|\mathbf{k}|)$. These constants can
possibly be complex, but appear in pairs with complex conjugates.
Substituting this approximation for $\hat{K}$ in equation
\eqref{eq:VpVqFourier} and taking the inverse Fourier transforms, we
obtain a system of {\color{red}coupled} partial differential equations with
possibly complex coefficients given by \beqs\label{eq:Vj} \becs
-\Delta V_{pj}+Q_{j}V_{pj}+P_{j}\Delta p(u)=&0 \qquad j=1\ldots m,\\
-\Delta V_{qj}+Q_{j}V_{qj}+P_{j}\Delta q(u)=&0 \qquad j=1\ldots m.
\eecs \eeqs where $V_{pj}$ and $V_{qj}$ are the inverse Fourier
transforms of
$\frac{P_j|\mathbf{k}|^2}{|\mathbf{k}|^2+Q_j}\hat{p}(\mathbf{k})$
 and $\frac{P_j|\mathbf{k}|^2}{|\mathbf{k}|^2+Q_j}\hat{q}(\mathbf{k})$ respectively
 for $j=1\ldots m$. Further, $V_p\approx\sum_{j}V_{pj}, V_q\approx\sum_{j}V_{qj}$.
 By defining $\varphi_{pj}=V_{pj}-P_{j}p(u)$ and $\varphi_{qj}=V_{qj}-P_{j}q(u)$ for $j=1\ldots m$,
 equation~\eqref{eq:Vj} can be rewritten as
\beqs\label{eq:varPhij} \becs
-\Delta \varphi_{pj}+Q_{j}\varphi_{pj}+P_{j}Q_{j}p(u)&=0 \qquad j=1\ldots m,\\
-\Delta \varphi_{qj}+Q_{j}\varphi_{qj}+P_{j}Q_j q(u)&=0 \qquad
j=1\ldots m. \eecs \eeqs The kernel energy, $E^{Ker}$, can now be
expressed in a local form in terms of the potentials $\varphi_{pj},
\varphi_{qj}$, or equivalently as a local saddle point problem:
\beqs
E^{Ker}(u)=\min_{\varphi_{pj}}\max_{\varphi_{qj}}\Big\{\sum_{j=1}^{m}\frac{1}{P_jQ_j}\int_{\Yp}\nabla\varphi_{pj} \cdot
\nabla\varphi_{qj}d\bfx+\frac{1}{P_j}\int_{\Yp}\varphi_{pj}\varphi_{qj}d\bfx \nonumber\\
+\int_{\Yp}\varphi_{qj}p(u)d\bfx+\int_{\Yp}\varphi_{pj}q(u)d\bfx+P_j\int_{\Yp}
p(u) q(u)d\bfx\Big\} .\eeqs
We note that variations with respect to
$\varphi_{pj}$ and $\varphi_{qj}$ return the Euler-Lagrange  equations
given by equation~\eqref{eq:varPhij}, and the saddle point problem correctly
represents, within the approximation~\eqref{eq:KerFourier}, the
kernel energy and its functional derivatives.

We now decompose the potential fields ($\varphi_{pj}, \varphi_{qj}$) into a
predictor ($\varphi_{{pj}_p}, \varphi_{{qj}_p}$) and a corrector
($\varphi_{{pj}_c}, \varphi_{{qj}_c}$), and expand the corrector
fields using a two-scale expansion given by \beqs\label{eq:varphiExp}  \becs
\varphi_{pj_c}(\bfx)=\varphi_{pj_c}^0(\bfx,\bfxtld)+\eta \varphi_{pj_c}^1(\bfx,\bfxtld)+\cdots, \\
\varphi_{qj_c}(\bfx)= \varphi_{qj_c}^0(\bfx,\bfxtld)+\eta \varphi_{qj_c}^1 (\bfx,\bfxtld)+\cdots.\\
\eecs \eeqs Following on similar lines as in
section~\ref{sec:homogenization}, we obtain  the following
expressions for $j=1\ldots m$ from the leading order terms of the expansion in equation~\eqref{eq:varphiExp}: \beqs \nabla_\bfxtld  \varphi_{pj_c}^0
(\bfx,\bfxtld)=0\qquad \aand\qquad
\nabla_\bfxtld  \varphi_{qj_c}^0 (\bfx,\bfxtld)=0,\\
\nabla_\bfxtld  \varphi_{pj_c}^1 (\bfx,\bfxtld)=0\qquad \aand\qquad
\nabla_\bfxtld  \varphi_{qj_c}^1 (\bfx,\bfxtld)=0. \eeqs Thus, the
corrector fields in their leading and first order are independent of
the fast variable representing the lattice length scale. The
governing equations for $\varphi_{pj_c}^0 (\bfx) \aand
\varphi_{qj_c}^0 (\bfx)$ are given by, \beqs \becs
-\Delta \varphi_{pj_c}^0+  Q_j \varphi_{pj_c}^0 + \xi_{p}(\bfFhat^\ast)\uczero =0&\oon\;\Yp,\\
-\Delta \varphi_{qj_c}^0+  Q_j \varphi_{qj_c}^0 +
\xi_{q}(\bfFhat^\ast)\uczero =0&\oon\;\Yp, \eecs \eeqs
where
$$\xi_{p}(\bfFhat^\ast)=\inttbar_{\bfFhat^\ast
U_0}p'(u_p(\bfFhat^\ast, \bfxtld)) d\bfxtld, \qquad
\xi_{q}(\bfFhat^\ast)=\inttbar_{\bfFhat^\ast
U_0}q'(u_p(\bfFhat^\ast, \bfxtld)) d\bfxtld.$$
 Further, the governing equations for $(\phiczero,\uczero)$ are given by \beqs
\becs
\Delta  \phiczero+  2 \alpha(\bfFhat^\ast) \uczero +b_c=0&\oon\;\Yp,\\
-\Delta \uczero + 2 \tilde{\gamma}(\bfFhat^\ast)\uczero +2
\alpha(\bfFhat^\ast) \phiczero +
\sum_{j=1}^m\big(\xi_{p}(\bfFhat^\ast)\varphi_{qj_c}^0+\xi_{q}(\bfFhat^\ast)\varphi_{pj_c}^0\big)=0 &\oon\;\Yp,\\
\eecs
 \eeqs
where
\beas
&&\tilde{\gamma}(\bfFhat^\ast)= \gamma (\bfFhat^\ast)
+ \half \sum_{j=1}^m\big(\chi_{pj}(\bfFhat^\ast)+\chi_{qj}(\bfFhat^\ast)+\psi_j(\bfFhat^\ast)\big),\\
&&\chi_{pj}(\bfFhat^\ast)=\inttbar_{\bfFhat^\ast U_0}p''(u_p(\bfFhat^\ast, \bfxtld)) \varphi_{qj_p}(\bfFhat^\ast,
\bfxtld) d\bfxtld, \\
&&\chi_{qj}(\bfFhat^\ast)=\inttbar_{\bfFhat^\ast U_0}q''(u_p(\bfFhat^\ast, \bfxtld)) \varphi_{pj_p}(\bfFhat^\ast,
\bfxtld) d\bfxtld,\\
&&\psi_j(\bfFhat^\ast)=\inttbar_{\bfFhat^\ast U_0}\Big[p''(u_p(\bfFhat^\ast, \bfxtld))q(u_p(\bfFhat^\ast, \bfxtld))
+2p'(u_p(\bfFhat^\ast, \bfxtld))q'(u_p(\bfFhat^\ast, \bfxtld))\\
&&\hspace{5cm}+p(u_p(\bfFhat^\ast, \bfxtld))q''(u_p(\bfFhat^\ast, \bfxtld))\Big]d\bfxtld.
\eeas

Finally, we comment that the results obtained with OFDFT as the model
theory are equally valid for the field formulations that describe
empirical interatomic potentials presented in \citet{IG2010}. We
note that the field formulation presented in \citet{IG2010} result
in a system of coupled linear partial differential equations which
represent a special case of the non-linear governing equations describing
OFDFT.

\section{Summary}\label{sec:conclusions}
The main idea behind the quasi-continuum reduction of field theories
is the coarse-graining of corrector fields in the formulation using
an unstructured finite-element triangulation. In this work we have
presented a formal mathematical justification that supports such a
coarse-graining, and places the quasi-continuum reduction of field
theories on a firm mathematical footing. In particular, we have
demonstrated using perturbation method and multiple scale analysis
that the corrector fields do not exhibit fine-scale (atomic-scale)
oscillations in the leading order, which allows for the
coarse-graining of these fields. Further, we have derived the
homogenized equations that govern the macroscopic far-field nature
of these corrector fields, and using Fourier analysis we have
estimated their far-field asymptotic behavior. In the case of
orbital-free density functional theory with TFW kinetic energy
functionals, the electronic fields comprising of the electrostatic
potential and electron density are found to exhibit an exponential
decay.

Using the computed asymptotic behavior of these corrector fields, we
have estimated the errors incurred in the computation of defect
energies using finite cell simulations. Although the electronic
fields exhibit an exponential decay, our analysis shows that
cell-sizes of the order of $10^3$ atoms are required for an accurate
computation of defect energies, which is in keeping with recent
cell-size studies conducted in \citet{GBO2007}. We note that in the
more accurate versions of density functional theory, like the
Kohn-Sham formulation, the decay in electronic fields is known to be
slower. Further, the asymptotic decay in elastic fields is much slower than electronic fields and this effect can become very significant for stronger defects like dislocations. This suggests that larger cell-sizes than those that are
typically used in electronic structure calculations ($\sim 100$
atoms) are needed for an accurate study of defects in materials.

\emph{A priori} estimates on the asymptotic behavior of corrector
fields from this work can be used to determine the optimal
coarse-graining rates for finite-element triangulations in the
quasi-continuum formulation of field theories, and presents itself
as a worthwhile future direction to pursue. Further, developing the
quasi-continuum reduction of Kohn-Sham density functional theory and an analysis of this
formulation is an important open problem, which is the focus of our
future work.

\section*{Acknowledgements}
We gratefully acknowledge the support of Air Force Office of
Scientific  Research under Grant No. FA9550-09-1-0240. The work of
V.G. also greatly benefited from the support of National Science
Foundation under Grant No. CMMI 0927478 and Army Research Office
under Grant No. W911NF-09-0292.

\end{document}